\documentclass[%
 reprint,
preprintnumbers,
nofootinbib,
amsmath,amssymb,
aps,
floatfix,
]{revtex4-2}

\usepackage{graphicx} 
\usepackage{dcolumn}  
\usepackage{colordvi}
\usepackage{color}
\usepackage{epstopdf}
\usepackage{amssymb}
\usepackage{url}
\graphicspath{{ps}}
\usepackage{hyperref}
\usepackage{tabularx}
\usepackage{amsmath}
\usepackage[most]{tcolorbox}
\tcbset{on line, 
        boxsep=4pt, left=0pt,right=0pt,top=0pt,bottom=0pt,
        colframe=white,  
        highlight math style={enhanced}
        }
\usepackage{pifont}
\usepackage{booktabs}
\usepackage{braket}
\usepackage{hepnames}
\usepackage{bm}
\usepackage{makecell}
\usepackage{appendix}
\usepackage{orcidlink}

\usepackage{mathrsfs}

\usepackage[noabbrev, nameinlink, capitalise]{cleveref}

\makeatletter
\renewcommand*{\thesection}{\arabic{section}}
\renewcommand*{\thesubsection}{\thesection.\arabic{subsection}}
\renewcommand*{\p@subsection}{}

\renewcommand*{\p@subsubsection}{}
\makeatother

\newcommand{\unit}[1]{\ensuremath{\, \mathrm{#1}}}
\newcommand\identity{1\kern-0.25em\text{l}}
\newcommand{\EOS}{\texttt{EOS}\xspace}
\newcommand{\pyhf}{\texttt{pyhf}\xspace}
\newcommand{\HistFactory}{\texttt{HistFactory}\xspace}

\begin{document}

\title{ \texorpdfstring{
    \quad\\[0.5cm] Constructing model-agnostic likelihoods, \\ a method for the reinterpretation of particle physics results
    }{Constructing model-agnostic likelihoods, a method for the reinterpretation of particle physics results}
    }

\author{Lorenz G\"artner \orcidlink{0000-0002-3643-4543}}
\email{lorenz.gaertner@physik.uni-muenchen.de}
\affiliation{Ludwig Maximilians University Munich, D-85748 Garching b.\ München, Germany}

\author{Nikolai Hartmann \orcidlink{0000-0003-0047-2908}}
\email{nikolai.hartmann@physik.uni-muenchen.de}
\affiliation{Ludwig Maximilians University Munich, D-85748 Garching b.\ München, Germany}

\author{Lukas Heinrich \orcidlink{0000-0002-4048-7584}}
\email{lukas.heinrich@cern.ch}
\affiliation{Technical University Munich, D-85748 Garching b.\ München, Germany}

\author{Malin Horstmann \orcidlink{0000-0002-4359-6364}}
\email{malin.horstmann@tum.de}
\affiliation{Technical University Munich, D-85748 Garching b.\ München, Germany}

\author{Thomas Kuhr \orcidlink{0000-0001-6251-8049}}
\email{thomas.kuhr@lmu.de}
\affiliation{Ludwig Maximilians University Munich, D-85748 Garching b.\ München, Germany}

\author{M\'{e}ril Reboud \orcidlink{0000-0001-6033-3606}}
\email{merilreboud@gmail.com}
\affiliation{Theoretische Physik 1, Naturwissenschaftliche Fakult\"at, Universit\"at Siegen, D-57068 Siegen, Germany}

\author{Slavomira Stefkova \orcidlink{0000-0003-2628-530X}}
\email{slavomira.stefkova@kit.edu}
\affiliation{Institut für Experimentelle Teilchenphysik, Karlsruhe Institute of Technology (KIT), D-76131 Karlsruhe, Germany}

\author{Danny van Dyk \orcidlink{0000-0002-7668-810X}}
\email{danny.van.dyk@gmail.com}
\affiliation{Institute for Particle Physics Phenomenology and Department of Physics, Durham University, Durham DH1 3LE, UK}


\preprint{IPPP/24/06}

\begin{abstract}
Experimental High Energy Physics has entered an era of precision measurements. However, measurements of many of the accessible processes assume that the final states' underlying kinematic distribution is the same as the Standard Model prediction. This assumption introduces an implicit model-dependency into the measurement, rendering the reinterpretation of the experimental analysis complicated without reanalysing the underlying data. We present a novel reweighting method in order to perform reinterpretation of particle physics measurements. It makes use of reweighting the Standard Model templates according to kinematic signal distributions of alternative theoretical models, prior to performing the statistical analysis. The generality of this method allows us to perform statistical inference in the space of theoretical parameters, assuming different kinematic distributions, according to a beyond Standard Model prediction. We implement our method as an extension to the \texttt{pyhf} software and interface it with the \texttt{EOS} software, which allows us to perform flavor physics phenomenology studies. Furthermore, we argue that, beyond the \texttt{pyhf} or \texttt{HistFactory} likelihood specification, only minimal information is necessary to make a likelihood model-agnostic and hence easily reinterpretable. We showcase that publishing such likelihoods is crucial for a full exploitation of experimental results.

\end{abstract}

\maketitle

\section{Introduction}
\label{sec:intro}
The results published using the data produced at high energy physics (HEP) experiments have large scientific potential beyond initial publication. To maximize the scientific impact of the data and corresponding results, facilitating reuse for combination and reinterpretation, should be made standard practice~\cite{Cranmer:2021urp}.

The importance of this is evident: Most analyses require underlying assumptions. These are, for example, theoretical distributions dictating signatures in the Monte Carlo (MC) data, which acts as a framework for constructing the analysis and provides a basis for comparison with the measured collider data. This also means, that a prior theoretical description has to be chosen, which typically corresponds to a Standard Model (SM) prediction. Therefore, the results obtained in the given analysis will be subject to a model dependency, which does not allow for simple reinterpretation in terms of alternative theories. 

The goal of reinterpretation efforts in HEP is to maximize the insight gained from existing collider data, which requires overcoming this model dependency. 
One can classify these reinterpretation efforts as follows~\cite{Bailey:2022tdz}:
\begin{itemize}
    \item \textit{Kinematic reinterpretation} or \textit{recasting}, which includes testing an alternative physics process with different kinematic distributions. Here, changes of efficiencies and acceptance regions need to be considered.
    \item \textit{Model updating}, which refines either theoretical predictions or experimental calibrations. This achieves an
    overall reduction of the uncertainties. Technically, this can be viewed as a subclass of kinematic reinterpretation.
    \item \textit{Combinations} of datasets and measurements across experiments. This is useful for reducing parameter uncertainties or for deriving global parameter constraints, where different decay channels have possibly different sensitivity to some parameters. For such combinations, it is necessary that the
    underlying model assumptions are mutually consistent.
\end{itemize}


Reinterpretation efforts have become a critical component of the research landscape~\cite{Cranmer:2021urp, Bailey:2022tdz}.
The main challenge remains the lack of public information on the analyses, that is, details available outside of the respective experimental collaborations. 
At the same time, the reinterpretation efforts are usually associated with a high computational cost due to the large number of theoretical models.
A comprehensive study of all theoretical models, ranging from MC production through analysis to statistical inference, is not feasible.

A review of common reinterpretation methods and tools can be found in references~\cite{Bailey:2022tdz,Cranmer:2021urp,Stark:2023ont,LHCReinterpretationForum:2020xtr}. 
Popular approaches can be classified as~\cite{LHCReinterpretationForum:2020xtr}:
\begin{itemize}
    \item \textit{Simulation based reinterpretation} (e.g. \texttt{CheckMate} \cite{Dercks:2016npn}, \texttt{MadAnalysis5} \cite{Conte:2012fm}, \texttt{RECAST}~\cite{Cranmer:2010hk}), where a full statistical analysis is performed on new MC samples produced according to an alternative theoretical model. 
    This requires access to details of the full analysis strategy, as well as the underlying collider data and potentially also individual MC samples. This information is usually not available outside experimental collaborations. In addition, this approach is very computationally resource-heavy as new MC samples must be produced and analysed for each alternative theory.
    \item \textit{Simplified model reinterpretation} (e.g. \texttt{SModelS} \cite{MahdiAltakach:2023bdn}), where one assumes that acceptances are not significantly affected by kinematic shape differences. Less information and computational resources are required, at the cost of approximations, which potentially lead to biases in the results (see \cref{sec:necessity}).
\end{itemize}

In this paper, we propose an alternative reinterpretation method based on the reweighting of simulated MC templates, as opposed to a reweighting of individual MC samples. The proposed method strikes a balance between the required information on analysis details and computational cost of bias-free reinterpretation.
Our work is an extension of the ``brief idea'' proposed in reference~\cite{cranmer_2017_1013926} and provides access to a reinterpretable likelihood function, directly parametrized in terms of any choice of theory parameters.

Reweighting is a standard practice in HEP commonly used for unfolding strategies, see e.g. reference~\cite{Pivk:2004ty}.
The \texttt{HAMMER} software provides an application of reweighting for the purpose of reinterpretation of experimental
measurements~\cite{Duell:2016maj,Bernlochner:2020tfi}; see also the interface to \texttt{RooFit}~\cite{GarciaPardinas:2020yrd}. 
At present, \texttt{HAMMER} allows for the reinterpretation of specifically implemented decays (mostly charged-current semileptonic $B$-meson decays) in terms of theoretical models of an effective field theory type by performing event-based reweighing. 
Our proposed reinterpretation method is more generally applicable, and it is not limited to any specific decay type or theoretical model.
Furthermore, the proposed method makes use of reweighting on the distribution level, rather than on the event level;
it does not require the full set of MC samples to perform the reinterpretation of the measurement;
and it is more efficient in terms of computational costs.

One prerequisite for making HEP measurements suitable for reinterpretation and/or combination is the distributability of statistical models. As discussed in reference~\cite{Cranmer:2021urp}, a general recommendation is to make sensibly parametrized likelihood functions publicly available. A standard for likelihood parametrization and preservation has been developed around the \pyhf software~\cite{pyhf,Heinrich:2021gyp} for statistical inference. The \pyhf software is an implementation of the \HistFactory model~\cite{histfactory}, which provides a general functional form for binned likelihoods. This fully parametrized binned likelihood is easily distributable in \texttt{JSON} format. In addition to the reinterpretation method proposed here, we also show that only a minimal amount of additional information allows for distributability of the \textit{reinterpretable} likelihood.

The paper is structured as follows.
In \cref{sec:reweighting method} we describe our novel reinterpretation method. We discuss the mathematical description and its applicability to unbinned and binned likelihoods, along with benefits and limitations. In \cref{sec:implementation} we describe the implementation of this method within the framework of common analysis tools. Finally, in \cref{sec:examples} we apply the reinterpretation method to two toy examples, which uses the model-agnostic framework of the Weak Effective Theory (WET). This effective theory covers all possible beyond Standard Model (BSM) theories that exclusively involve new particles or interactions at or above the scale of electroweak symmetry breaking.
We also compare our newly developed method to a more simple approach, commonly used to reinterpret HEP results.

\section{Reweighting method}
\label{sec:reweighting method}
The reinterpretation method described here is based on updating the distributions of the observable variables, given changes in the underlying kinematic distribution.

The probability density function (PDF) of reconstructed events $p(x)$ results from folding the PDF of a theoretical kinematic prediction $p(z)$ with the conditional distribution $p(x|z)$ and the indicator function $\identity_\varepsilon(x)$,
\begin{equation}
    p(x) = \frac{1}{\varepsilon}\int dz ~ \identity_\varepsilon(x) ~ p(x|z) ~ p(z).
\end{equation}
Here, the reconstruction variable $x$ represents one or multiple observable variables and the kinematic variable $z$ represents one or multiple kinematic degrees of freedom (d.o.f.).
The function $\identity_\varepsilon(x)$ models the selection criteria for a reconstructed event and $p(x|z)$ is the conditional probability of measuring a reconstructed configuration $x$, given an underlying particle configuration $z$. 
The overall reconstruction efficiency $\varepsilon$ acts as a normalization factor for the PDF $p(x)$. 
The PDF $p(z)$ corresponds to the normalized kinematic distribution of a theoretical prediction $\sigma(z)$,
\begin{equation}
    p(z) = \frac{\sigma(z)}{\sigma}.
\end{equation}

The number density of expected events, given a total integrated luminosity $L$ is $n(x) = L ~ \sigma ~\varepsilon ~p(x)$ and further reads
\begin{equation}
    n(x) = L \int dz ~ \varepsilon(x|z) ~ \sigma(z) = \int dz ~ n(x,z),
\end{equation}
where we combine both reconstruction and selection into $\varepsilon(x|z) = \identity_\varepsilon(x) ~ p(x|z)$, and where $n(x,z) = L ~ \varepsilon(x|z) ~ \sigma(z)$ can be thought of as a joint number density, similar to the joint PDF $p(x,z) = p(x|z) ~ p(z)$.

The reinterpretation task involves determining the number density $n_1(x)$ of an \textit{alternative} theoretical prediction $\sigma_1(z)$. This can be obtained by reweighting the joint number density $n_0(x,z)$ according to the kinematic \textit{null} distribution $\sigma_0(z)$, via
\begin{equation}
    \begin{aligned}
    n_1(x)  &= L \int dz ~ \varepsilon(x|z) ~ \sigma_1(z) \\
            &= L \int dz ~ \varepsilon(x|z) ~ \sigma_0(z) ~ \frac{\sigma_1(z)}{\sigma_0(z)}\\
            &= \int dz ~ n_0(x,z) ~ w(z).
    \end{aligned}
    \label{eq:reweight}
\end{equation}

The weight factor $w(z)$ is simply the ratio of the theoretically predicted alternative kinematic distribution to the null distribution. 

This reweighting process solely requires the knowledge of the joint null number density $n_0(x,z)$. Together with the weight factor, this is enough to predict the number density according to an alternative theory.

\subsection{Discrete reweighting}
\label{sec:reweighting-method-discrete}
In practical applications, the continuous joint number density is typically not analytically obtainable and requires estimation through MC simulations.
To address this, one can discretize the reweighting method by representing the joint number density as a multidimensional matrix $ n_{xz}$ in bins of $x \times z$. This is done alongside the binning of the theoretically predicted distribution $\sigma_{z}$ and weight factor $w_{z}$ in the kinematic d.o.f. $z$. Consequently, the discrete joint number density has dimension $\dim(x) \times \dim(z)$. 

The discrete joint number density can be obtained by integrating the continuous joint number density over each $x \times z$ bin,
\begin{equation}
    n_{xz} = \int_{\text{bin } x} \int_{\text{bin }  z} dx' ~ dz' ~ n(x',z'),
\end{equation}
where the integral boundaries are the bin boundaries of each $x$ and $z$ bin, respectively. The binned weights are given by 
\begin{equation}
    w_z = \frac{\sigma_{1,z}}{\sigma_{0,z}} = \frac{\int_{\text{bin }  z} dz' ~ \sigma_1(z')}{\int_{\text{bin }  z} dz' ~ \sigma_0(z')} ~ .
    \label{eq:binned-weights}
\end{equation}

The reweighting step of \cref{eq:reweight} becomes
\begin{equation}
    n_{1,x} = \sum_{z} ~ n_{0,xz} ~ w_{z}
    \label{eq:reweight_discrete}
\end{equation}

We see an advantage in using this discrete approach because of lower computations costs.
The price for this simplification is a loss of accuracy due to the binning in the kinematic d.o.f. $z$. However, this loss is controllable by increasing the number of bins. 
The binning should be chosen such that the joint number density and weight function factorize approximately in each bin.
In principle, an arbitrarily fine binning can be chosen such that the uncertainty due to this loss is negligible compared to other sources of uncertainty (provided enough MC samples are available; see \cref{app:kinematic-binning}).

Crucially, only a fixed set of samples from the joint PDF $p_0(x,z)$, based on the null prediction, is required, and no new samples need to be produced for the reinterpretation.
Therefore, publishing the (binned) joint null number density and knowledge about the underlying kinematic null distribution is sufficient to perform the reinterpretation of a given measurement.

\subsection{Limitation}
The proposed reweighting approach is a light-weight and accurate way of obtaining new signal templates, given a joint number density and the kinematic null distribution. Still, it does have one main limitation:
if the phase space contains regions with $w(z) \gg 1$, the effective MC sample size decreases.
Put differently, we assign large weights to regions that are sparsely populated with MC samples obtained from the null distribution.
Further, if the null distribution lacks support in a region of phase space, $\sigma_0(z) \to 0$, it can happen that $w(z) \to \infty$ when reweighting to an alternative distribution. In this case, we have no MC samples in the given region. The only solution that we see is reanalysing new samples, produced according to the alternative distribution.

\section{Implementation and likelihood construction}
\label{sec:implementation}
Using the reweighting method, we can construct likelihood functions for particle physics analyses, directly parametrized in terms of theory parameters. Even though the reweighting method is independent of any likelihood formalism, we showcase our method in terms of the \HistFactory formalism~\cite{histfactory} as a baseline statistical model.

To build a global likelihood or posterior for a given measurement, including theoretical constraints or priors, we split the likelihood into three parts. The total likelihood is a combination of a data likelihood, $L_\text{data}$, the experimental constraint, $C_\text{ex}$, and the theory constraint, $C_\text{th}$,
\begin{equation}
    L = L_\text{data} \cdot C_\text{ex} \cdot C_\text{th}.
    \label{eq:likelihood}
\end{equation}
The data likelihood is constructed as a product of the Poisson probabilities of experimentally obtaining $\boldsymbol{n}$ events, when $\boldsymbol{\nu}$ are expected from MC simulation, 
\begin{equation}
    L_\text{data}(\boldsymbol{n} | \boldsymbol{\eta}, \boldsymbol{\chi})=\prod_{c \in \text {channels }} \prod_{b \in \text {bins}} \operatorname{Pois}\left(n_{c b} | \nu_{c b}(\boldsymbol{\eta}, \boldsymbol{\chi})\right).
\end{equation}
\textit{Channels} represent disjoint binned distributions, for example signal and control channels. \textit{Bins} correspond to the histogram bins. The expected bin counts $\boldsymbol{\nu}$ are a function of unconstrained, $\boldsymbol{\eta}$, and constrained, $\boldsymbol{\chi}$, parameters.

The experimental constraint consists of constraint terms or priors for experimental nuisance parameters $\boldsymbol{\chi}_\text{ex} \subset \boldsymbol{\chi}$, including all experimental systematic uncertainties,
\begin{equation}
    C_\text{ex}(\boldsymbol{a} | \boldsymbol{\chi}_\text{ex})= \prod_{\chi \in \boldsymbol{\chi}_\text{ex}} c_\chi\left(a_\chi | \chi\right).
\end{equation}
In the frequentist language, constraints, $c_\chi(a_\chi | \chi)$, are obtained from auxiliary measurements with corresponding auxiliary data $\boldsymbol{a}$. This is the frequentist parallel of a \textit{prior distribution}.

The theory constraint consists of constraint terms or priors for theoretical parameters, $\boldsymbol{\chi}_\text{th} \subset \boldsymbol{\chi}$,
\begin{equation}
    C_\text{th}(\boldsymbol{a} | \boldsymbol{\chi}_\text{th})= \prod_{\chi \in \boldsymbol{\chi}_\text{th}} c_\chi\left(a_\chi | \chi\right).
\end{equation}

\subsection{Implementation of the reweighting method}
\label{sec:implementation-reweighting}
To obtain the data likelihood $L_\text{data}$ for any theoretical model, one needs to calculate the event rates of the corresponding signal template. This requires an implementation of \cref{eq:reweight_discrete}.

To achieve this, we work with \pyhf~\cite{pyhf,Heinrich:2021gyp}, which is an implementation of the \HistFactory model~\cite{histfactory}.  Here, a likelihood is constructed by specifying the bin content of all contributing signal and background processes, as usually obtained from MC simulation, and the data measured in an experiment.

Furthermore, to implement uncertainties, one needs to specify the properties of a set of \textit{modifiers}. The modifier settings include the event rate modifications according to each type of uncertainty at the $1\sigma$ level and the corresponding constraint type for the modifier parameter. The event rates for each channel and bin are calculated as
\begin{equation}
    \begin{aligned}
      \nu_{c b}(\boldsymbol{\eta}, \boldsymbol{\chi})=\sum_{s \in \text {samples }} &\prod_{\kappa \in \boldsymbol{\kappa}} \kappa_{s c b}(\boldsymbol{\eta}, \boldsymbol{\chi}) \times \\
      &\left( \nu_{s c b}^0(\boldsymbol{\eta}, \boldsymbol{\chi})+\sum_{\Delta \in \boldsymbol{\Delta}} \Delta_{s c b}(\boldsymbol{\eta}, \boldsymbol{\chi}) \right),
    \end{aligned}
\label{eq:counts}
\end{equation}
where \textit{samples} are used to separate physics processes, $\boldsymbol{\nu}^0$ are the nominal event rates (determined from MC), $\boldsymbol{\kappa}$ and $\boldsymbol{\Delta}$ are the full set of multiplicative and additive modifiers, respectively.

The implementation of the reweighting method (\cref{sec:reweighting method}) requires extending the \pyhf codebase.\footnote{%
    We plan to contribute our modifications to the \pyhf codebase.
}
The method is a prescription on the change of event rates. Therefore, a multiplicative modifier can be used to apply these changes. We extend \pyhf by a \textit{custom modifier}, which calculates the modifications to the nominal event rates according to the procedure described in \cref{eq:reweight_discrete}. This custom modifier is a function of the underlying theory parameters of the alternative kinematic distribution.

The theory constraint $C_\text{th}$ contains all constraint terms of these underlying theory parameters, which can be correlated in general. Per definition, modifier parameters in \pyhf are treated as uncorrelated. To correctly include correlated parameter constraints in our statistical model, we decorrelate the theory parameters using principal component analysis (see \cref{app:svd}). We then assign one normally constrained \pyhf modifier parameter to each of these decorrelated parameters. 

\section{Example application}
\label{sec:examples}
A central aim of this work is to motivate the experimental HEP community to make use of the proposed method.
This will, in turn, enable subsequent model-agnostic phenomenological analyses of the experimental results.
Here, we detail the full reinterpretation of an analysis result of two toy examples from low-energy flavour physics.

In general, one distinguishes between two datasets for a given analysis: the \textit{real data}, measured by a collider experiment and a set of simulated signal and background \textit{MC data}, produced as a means of comparing against the measured collider data.

In this toy example, we do not use any experimental data, but two datasets of simulated samples, where one dataset acts as \textit{real} data. The \textit{MC data} is produced according to the SM prediction. The \textit{real data} is produced by assuming that BSM physics affects the example process. 
To simulate detector and other analysis selection effects,
observables in both datasets are smeared according to estimated detector resolutions, and event yields are scaled by an efficiency map.

By comparing the produced datasets, using either Bayesian or frequentist methods, we aim to recover the chosen BSM parameters, starting from the SM. This is possible only, because we made the likelihood a function of the theory parameters and a shape change in the kinematic distribution due to BSM physics can be directly taken into account.

As a general result, we want to compute the posterior distribution in the space of theory parameters, given the two simulated datasets and prior parameter constraints. To obtain a posterior from the likelihoods of the form shown in \cref{eq:likelihood}, we use the \texttt{bayesian pyhf}~\cite{Feickert:2023hhr}, an extension to \pyhf. 
The posterior is obtained by Markov chain Monte Carlo (MCMC) sampling from the total likelihood, following Bayes' theorem for auxiliary data $\boldsymbol{a}$ and observations $\boldsymbol{n}$,
\begin{equation}
    \begin{aligned}
        p\left( \boldsymbol{\eta}, \boldsymbol{\chi} \vert \boldsymbol{n}, \boldsymbol{a} \right) \propto L_\text{data}\left( \boldsymbol{n} \vert \boldsymbol{\eta}, \boldsymbol{\chi} \right) p\left( \boldsymbol{\chi} | \boldsymbol{a} \right) p\left( \boldsymbol{\eta} \right).
    \end{aligned}
    \label{eq:BayesTheorem}
\end{equation}
The experimental and theoretical constraints are represented by the constraint priors $p\left( \boldsymbol{\chi} | \boldsymbol{a} \right)$, and $p\left( \boldsymbol{\eta} \right)$ contains the priors for the unconstrained parameters $\boldsymbol{\eta}$ as detailed in reference~\cite{Feickert:2023hhr}.

\subsection{\texorpdfstring{$B \to K \nu \bar \nu$}{B -> K nu nu}}
\label{sec:knunu}
The recent measurements of the total rate of $B \to K \nu \bar \nu$ decays by the Belle II collaboration~\cite{Belle-II:2021rof,Belle-II:2023esi} hint at an excess of signal events compared to the SM expectation. This has triggered a substantial interest in the HEP phenomenology community to interpret this excess as a sign of BSM physics and to extract the corresponding model parameters~\cite{Fridell:2023ssf,Altmannshofer:2023hkn, Gabrielli:2024wys}.
In this subsection, we study the performance of our proposed approach at the hand of simulated $B\to K\nu\bar\nu$ data.

\subsubsection{Weak Effective Theory parametrization}
\label{sec:knunu-theory}
While we cannot achieve a general model-independent theoretical description of the $B\to K \nu \bar\nu$ decay, it is nevertheless possible to capture the effects of a large number of BSM theories under mild assumptions, as mentioned previously.
Here, we assume that potential new BSM particles and force carriers have masses at or above the scale of electroweak symmetry breaking.
In this scenario, it is useful to work within an effective quantum field theory that describes both the SM and the potential BSM effects using a common set of parameters; this effective field theory is commonly known as the Weak Effective Theory (WET)~\cite{Aebischer:2017gaw,Jenkins:2017jig,Jenkins:2017dyc}.

For the description of $b\to s\nu\bar\nu$ transitions, it suffices to discuss the $sb\nu\nu$ \emph{sector} of the WET.
It is spanned by a subset of local operators of mass-dimension six, which is closed under the renormalization group~\cite{Aebischer:2017ugx}.
Since the mass of the initial on-shell $B$ meson limits the maximum momentum transfer in this process, the matrix elements of
operators with mass dimension eight or above are suppressed by at least a factor of $M_B^2 / M_W^2 \simeq 0.004$, which
are hence commonly neglected in these types of analyses.
The corresponding Lagrangian density for the $sb\nu\nu$ sector reads~\cite{Felkl:2021uxi}
\begin{equation}
    \mathcal{L}^\text{WET}= -\frac{4 G_\text{F}}{\sqrt{2}} \frac{\alpha}{2 \pi} V_{t s}^* V_{t b}
    \sum_i C_i(\mu_b) O_i + \text{h.c.}\,,
    \label{eq:lagrangian}
\end{equation}
with $G_F$ the Fermi constant, $\alpha$ the fine structure constant, and $V$ the Cabibbo-Kobayashi-Maskawa quark mixing matrix, respectively.
The separation scale is chosen to be $\mu_b = 4.2~\unit{GeV}$. Matrix elements of the operators $\mathcal{O}_i$ describe
the dynamics of the process at energies below $\mu_b$, while the dynamics at energies above $\mu_b$ are encoded in the
(generally complex-valued) Wilson coefficients $C_i(\mu_b)$ in the
modified minimal subtraction ($\overline{\text{MS}}$) scheme. This enables a simultaneous description of SM-like and
BSM-like dynamics in $b\to s\nu\bar\nu$ processes, as long as all BSM effects occur at scales larger than $\mu_b$;
the different dynamics result simply in different values of the Wilson coefficients.

Assuming massless neutrinos, the full set of dimension-six operators
is given by~\cite{Felkl:2021uxi},
\begin{equation}
\begin{aligned}
\mathcal{O}_{\mathrm{VL}} &=\left(\overline{\nu_L} \gamma_\mu \nu_L\right)\left(\overline{s_L} \gamma^\mu b_L\right) \\
\mathcal{O}_{\mathrm{VR}} &=\left(\overline{\nu_L} \gamma_\mu \nu_L\right)\left(\overline{s_R} \gamma^\mu b_R\right) \\
\mathcal{O}_{\mathrm{SL}} &=\left(\overline{\nu_L^c} \nu_L\right)\left(\overline{s_R} b_L\right) \\
\mathcal{O}_{\mathrm{SR}} &=\left(\overline{\nu_L^c} \nu_L\right)\left(\overline{s_L} b_R\right) \\
\mathcal{O}_{\mathrm{TL}} &=\left(\overline{\nu_L^c} \sigma_{\mu \nu} \nu_L\right)\left(\overline{s_R} \sigma^{\mu \nu} b_L\right),
\end{aligned}
\label{eq:operators}
\end{equation}
with
\begin{equation}
\begin{aligned}
    \overline{\nu_L}
        & \equiv \nu_L^\dagger \gamma^0 \,, \\
    \nu_L^c
        & \equiv C \overline{\nu_L}^T   \,,
\end{aligned}
\end{equation}
and where $C=i\gamma^2\gamma^0$ is the charge conjugation operator. 
In the above, the subscripts $\text{V},\text{S},\text{T}$ represent vectorial, scalar, and tensorial operators, respectively;
$\nu_{L/R}$ represent left- or right-handed neutrino fields;
and $q_{L/R}$ represent left- or right-handed quark fields.
The spin structure of the operators is expressed in terms of the Dirac matrices $\gamma^\mu$ and their commutator $\sigma^{\mu\nu} \equiv \frac{i}{2} [\gamma^\mu, \gamma^\nu]$.
The operators are defined as sums over the neutrino flavors, since this is a property that we cannot determine experimentally.
If one assumes the existence of only left-handed massless neutrinos, all operators
except $\text{VL}$ and $\text{VR}$ vanish.
The SM point in the parameter space of the WET Wilson coefficients reads
\begin{equation}
\label{eq:WET-SM-point}
\begin{aligned}
    C_\text{VL} & \simeq 6.6\,,                   &
    C_i         & = 0 \quad \forall \quad i \neq \text{VL}\,.
\end{aligned}
\end{equation}

\begin{figure}[t]
    \centering
    \includegraphics[width=\linewidth]{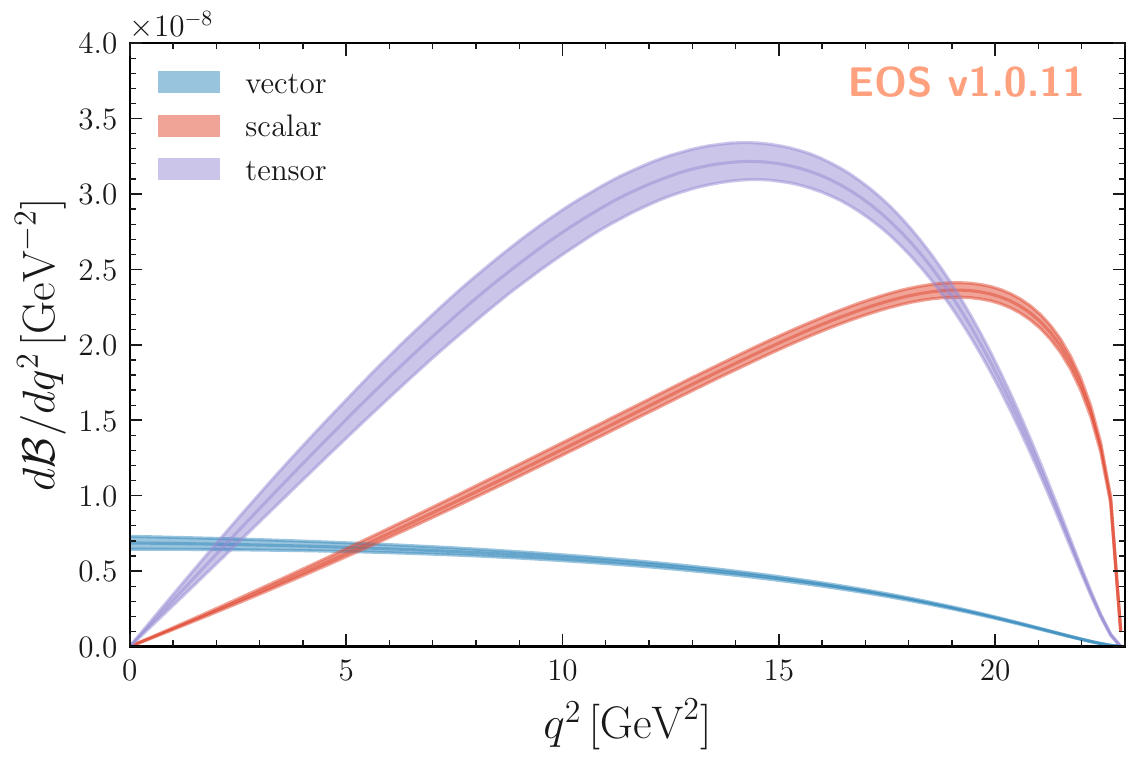}
    \caption{%
        Illustration of the variety of shapes of the $B \to K \nu \bar \nu$ decay rate due to purely vectorial, scalar, or tensorial interactions.
        Each curve corresponds to setting a single non-zero Wilson Coefficient in \cref{eq:width} to unity while keeping
        all other coefficients at zero.
    }
    \label{fig:knunu-theory}
\end{figure}
Presently, the only measured observable is the differential decay rate for $B\to K \nu \bar{\nu}$, which we simulate in
this example.
Since the $B$ meson is a pseudoscalar, the decay is isotropic in the rest frame of the $B$ meson and hence the only kinematically free variable is the squared dineutrino invariant mass, $q^2 = (p_B - p_K)^2 = (p_\nu + p_{\bar \nu})^2$. The differential decay rate is given by~\cite{Gratrex:2015hna,Felkl:2021uxi}
\begin{equation}
    \begin{aligned}
      \frac{d \Gamma}{d q^{2}}
      & =
      3
      \left(\frac{4 G_\text{F}}{\sqrt{2}} \frac{\alpha}{2 \pi} \right)^2 \left|V_{t s}^* V_{t b}\right|^2
      \frac{\sqrt{\lambda_{B K}} q^{2}}{(4 \pi)^{3} M_{B}^{3}}\\
      &\times\left[\frac{\lambda_{B K}}{24 q^{2}}\left|f_{+}(q^2)\right|^{2}\left|C_{\mathrm{VL}}+C_{\mathrm{VR}}\right|^{2}\right.\\
      &\phantom{\times}+\frac{\left(M_{B}^{2}-M_{K}^{2}\right)^{2}}{8\left(m_{b}-m_{s}\right)^{2}}\left|f_{0}(q^2)\right|^{2}\left|C_{\mathrm{SL}}+C_{\mathrm{SR}}\right|^{2} \\
      &\phantom{\times}\left.+\frac{2 \lambda_{B K}}{3\left(M_{B}+M_{K}\right)^{2}}\left|f_{T}(q^2)\right|^{2}\left|C_{\mathrm{TL}}\right|^{2}\right],
    \end{aligned}
\label{eq:width}
\end{equation}
where $M_B, M_K$ are the masses of the $B$ meson and the kaon, respectively, $m_b, m_s$ are the masses of the $b$ and $s$ quarks in the $\overline{\text{MS}}$ scheme, respectively, and $\lambda_{B K} \equiv \lambda(M_B^2, M_K^2, q^2)$ is the  K\"all\'en function. \\
In order to highlight the individual contributions of vectorial, scalar and tensorial terms from \cref{eq:width} to the differential decay rate, an illustration where individual left-handed Wilson coefficients are set to unity, is shown in \cref{fig:knunu-theory}.

As can be inferred from \cref{eq:width}, the decay is only sensitive to the magnitude of three linear combinations
of Wilson coefficients
\begin{equation}
    |C_{VL}+C_{VR}|, ~ |C_{SL}+C_{SR}|, ~ |C_{TL}|.
    \label{eq:BToKnunu-wc-sensitivity}
\end{equation}
The hadronic matrix elements of their operators are described by three independent hadronic form factors commonly known as $f_{+}(q^2)$, $f_{0}(q^2)$ and $f_{T}(q^2)$, which are functions of $q^2$.
In this work, the form factors are parametrized following the BCL parametrization~\cite{Bourrely:2008},
which is truncated at order $K=2$.
The values for the corresponding $8$ hadronic parameters are extracted from a joint theoretical
prior PDF comprised of the 2021 lattice world average based on results by the Fermilab/MILC and HPQCD
collaborations~\cite{FlavourLatticeAveragingGroupFLAG:2021npn},
and recent results by the HPQCD collaboration~\cite{Parrott:2022rgu}.
Correlations between the hadronic parameters are taken into account through their respective covariance matrices and implemented as discussed in \cref{sec:implementation-reweighting}.

The Belle II experiment, which found the first evidence for this decay, observes more events than expected in the SM.
The ratio of observed to expected events is ${4.6 \pm 1.3}$~\cite{Belle-II:2023esi}.
For latter use, we define a benchmark point in the space of Wilson coefficients that roughly reproduce the measured branching fraction, after correcting for the efficiency.
Assuming all Wilson coefficients to be real, it reads
\begin{equation}
    \label{eq:WET-BSM-benchmark}
    \begin{aligned}
        |C_{VL}| & = 10\,, \qquad &|C_{VR}| =  4\,, \\
        |C_{SL}| & =  3\,, \qquad &|C_{SR}| =  1\,, \\
        |C_{TL}| & =  1\,.
    \end{aligned}
\end{equation}
As shown in \cref{eq:width}, the decay rate of $B\to K\nu\bar\nu$ is only sensitive to three linear
combinations of Wilson coefficients shown in \cref{eq:BToKnunu-wc-sensitivity}.
The projection of our benchmark point onto this subspace reads
\begin{equation}
    \label{eq:WET-BSM-benchmark-Knunu}
    \begin{aligned}
        |C_{VL}+C_{VR}| & = 14\,, \\
        |C_{SL}+C_{SR}| & =  4\,, \\
        |C_{TL}|        & =  1\,.
    \end{aligned}
\end{equation}

\subsubsection{Datasets}
\label{sec:knunu-data}

\begin{table}[t]
    \centering
    \begin{tabular}{ cccc } 
    \hline
    Luminosity & $B \overline{\kern -0.18em B} $ events & \makecell{MC generated \textbackslash \\ reconstructed} &\makecell{Data generated \textbackslash \\ reconstructed} \\
    \hline
    $362~\text{fb}^{-1}$ & $\sim 3.87 \times 10^8$ & \makecell{$1.86\times 10^3$ \\ 241} & \makecell{ $1.05\times 10^4$ \\ $1.14\times 10^3$} \\ 
    $50~\text{ab}^{-1}$   & $\sim 5.35 \times 10^9$ & \makecell{$2.57\times 10^5$ \\ $3.21\times 10^4$} & \makecell{$1.45 \times 10^6$ \\ $1.58 \times 10^5$}\\ 
    \hline
    \end{tabular}
    \caption{The number of $B \to K \nu \bar \nu$ samples produced for this study, corresponding to an equivalent of $362~\text{fb}^{-1}$ and $50~\text{ab}^{-1}$ integrated luminosity at the SuperKEKB collider. \textit{Generated} and \textit{reconstructed} samples correspond to the numbers prior and post efficiency correction.}
    \label{tab:samples}
\end{table}

To make this example as realistic as possible, we design our setup similar to what has been done in the Belle II analysis.

The MC data is produced according to the SM prediction (\textit{null} hypothesis) of the differential branching ratio $d\mathcal{B}(B \to K \nu \bar \nu)/dq^2$, where the Wilson coefficient are set to the values in \cref{eq:WET-SM-point}.
The number of samples produced is equivalent to the number of signal events seen or expected for a given integrated luminosity.
We produce samples for $362~\text{fb}^{-1}$ integrated luminosity, which corresponds to the total collider data used in reference~\cite{Belle-II:2023esi}, and for $50~\text{ab}^{-1}$ integrated luminosity, corresponding to the total target luminosity of the Belle II experiment~\cite{Belle-II:2018jsg}, respectively. 
We multiply the estimated number of $B \overline{\kern -0.18em B}$ events with the SM branching fraction, $BR(B \to K \nu \bar \nu) \approx 4.81 \times 10^{-6}$~\cite{EOSAuthors:2021xpv,EOS:v1.0.11} to get a rough estimate for the number of MC samples to produce, prior to the efficiency correction (see below).

The real data is produced according to a BSM prediction (\textit{alternative} hypothesis).
Following the observations of more events than predicted in the SM~\cite{Belle-II:2023esi},
we use the previously defined benchmark point in \cref{eq:WET-BSM-benchmark}. We multiply the estimated number of $B \overline{\kern -0.18em B}$ events with the BSM branching fraction, $BR(B \to K \nu \bar \nu) \approx 2.71 \times 10^{-5}$~\cite{EOSAuthors:2021xpv,EOS:v1.0.11} to get a rough estimate for the number of data samples to produce, prior to the efficiency correction (see below).

We list the number of produced samples in \cref{tab:samples}.
Unless stated otherwise, all numerical values, figures, and tables provided in the following are obtained
from studies that assume the $50~\text{ab}^{-1}$ sample size.
We produce samples of the decay's probability distribution for both the null and the alternative hypothesis
using the \EOS software in version 1.0.11~\cite{EOS:v1.0.11}.

To simulate the detector resolution, we shift the $q^2$ value of each sample by a value drawn from a normal distribution of width $1~\text{GeV}^2$. This roughly corresponds to the Belle II detector resolution. 

Furthermore, we apply an efficiency map to the samples according to the function 
\begin{equation}
    \varepsilon(q^2) = 0.4 ~ \exp\left(- 5 ~ q^2/M_B^2\right),
    \label{eq:eff-knunu}
\end{equation}
which mimics the efficiency obtained in reference~\cite{Belle-II:2023esi}.

\begin{figure}[ht]
    \centering
    \includegraphics[width=\linewidth]{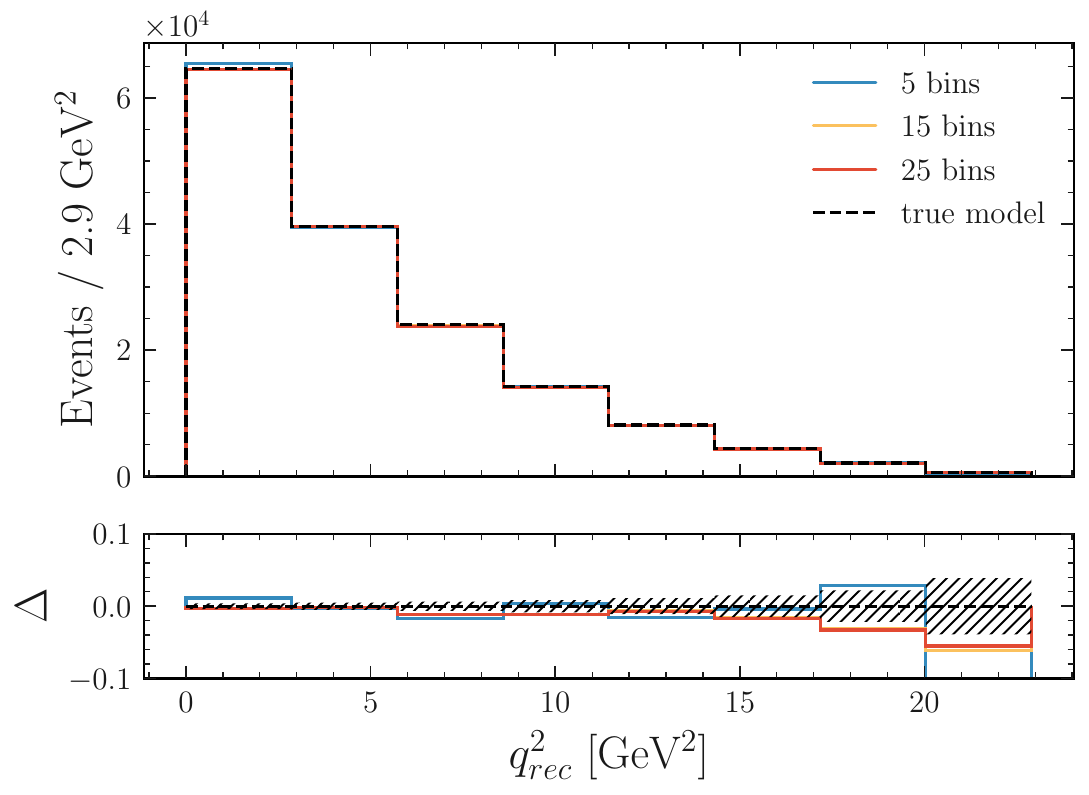}
    \caption{The null histogram yields, reweighted to the benchmark point in \cref{eq:WET-BSM-benchmark}, as a function of kinematic bins (red). The histogram yields of the true model, sampled from the probability density function at the benchmark point (black). The bottom plot shows the relative difference $\Delta$ of the reweighed models to the true model. The statistical uncertainty of the true model is shown as the hatched region.}
    \label{fig:knunu-binning}
\end{figure}

The reconstruction variable is chosen to be the reconstructed momentum transfer $q_{rec}^2$, obtained from the kinematic variable $q^2$, by applying detector and efficiency correction. 
The binnings of our kinematic and our reconstruction variables differ:
\begin{itemize}
\item For the reconstruction variable,
we need to strike a balance between the number of events in each bin (to ensure sufficient statistical power) and the number of bins to ensure sensitivity toward differences in the shape of the $q^2$ distribution.
We choose 8 equally spaced bins for the reconstruction variable.
\item For the kinematic variable, we
determine the number of bins as follows.
We study the convergence of the expected yields from the reweighted model, as we increase the number of kinematic bins.
For this study, we remove the detector resolution smearing.
This is done for 100 randomly chosen theoretical models (see \cref{app:kinematic-binning} for further details).
These models correspond to normally-distributed variations $\sim \mathcal{N}(0, 10)$ of the WET parameters with respect to the SM parameter point.
We aim to ensure convergence at the level of $1\%$ accuracy. We find that using 24 equally spaced bins for the kinematic variable ensures this aim. 
\Cref{fig:knunu-binning} illustrates this procedure for the benchmark point in \cref{eq:WET-BSM-benchmark}.
\end{itemize}

\begin{figure}[ht]
    \centering
    \includegraphics[width=\linewidth]{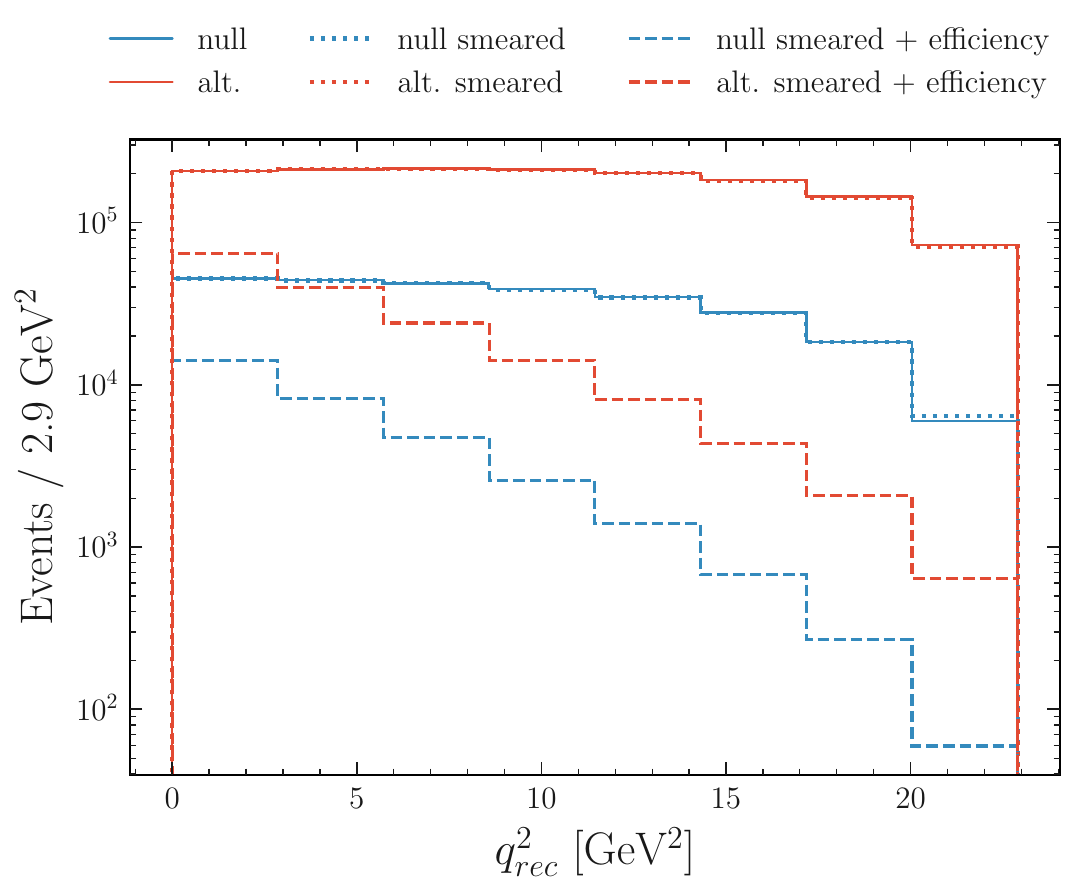}
    \caption{Both the null/SM (blue) and alternative/BSM (red) $B \to K \nu \bar \nu$ datasets, according to the pure theoretical prediction, after detector resolution smearing and efficiency correction.}
    \label{fig:knunu-data-sets}
\end{figure}

Both datasets, according to the null (SM) and alternative (BSM) hypothesis, and their corresponding changes after detector resolution smearing and efficiency correction are shown in \cref{fig:knunu-data-sets}.

\begin{figure}[ht]
    \centering
    \includegraphics[width=\linewidth]{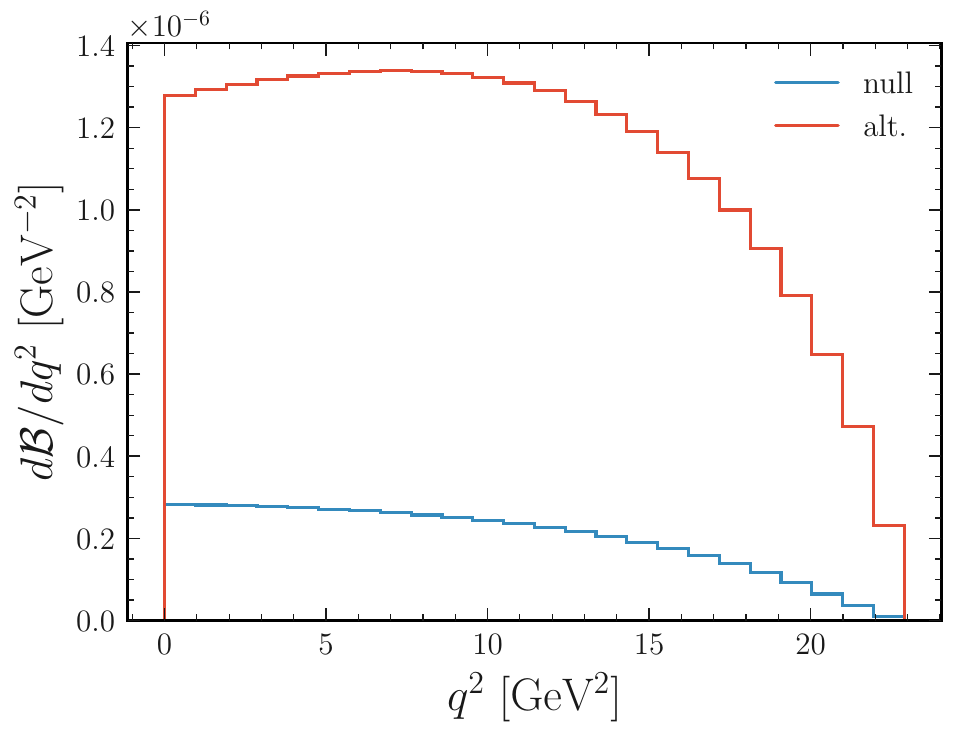}
    \caption{The bin-integrated null/SM (blue) and alternative/BSM (red) predictions for the differential branching ratio $d\mathcal{B}(B \to K \nu \bar \nu)/dq^2$.}
    \label{fig:knunu-dist}
\end{figure}

\begin{figure}[ht]
    \centering
    \includegraphics[width=\linewidth]{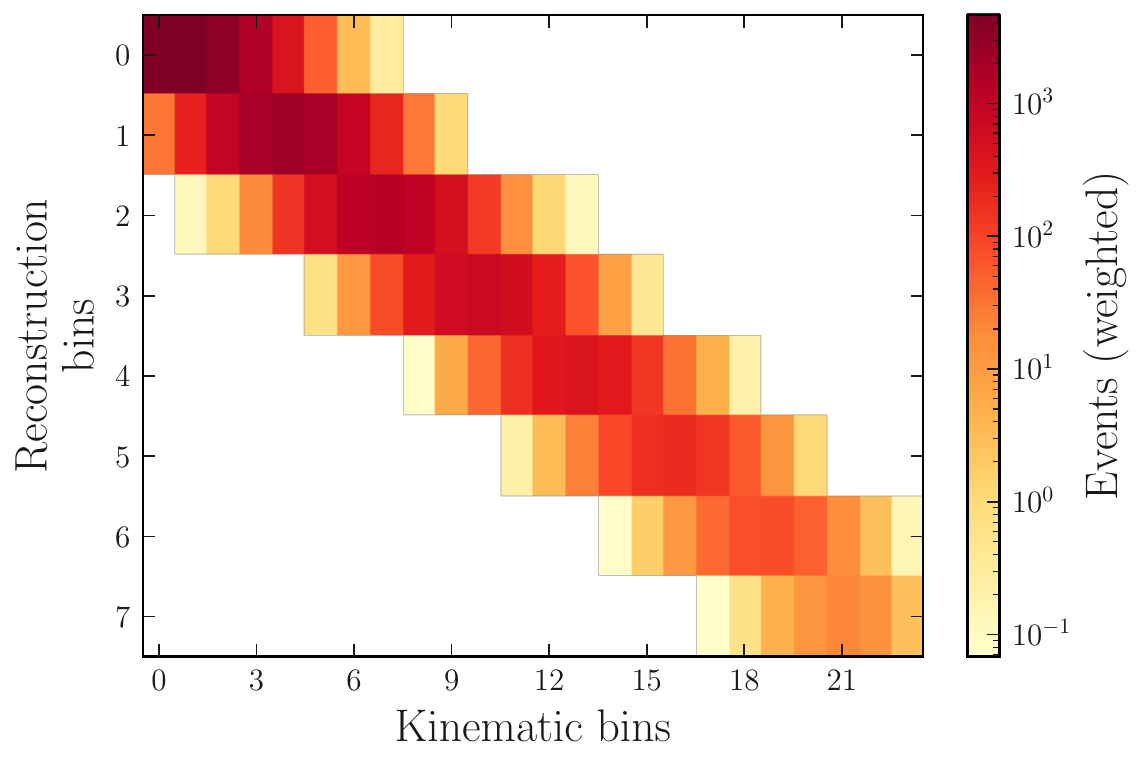}
    \caption{The null joint number density, where we see the 8 bins of the reconstruction variable on the vertical axis and the 24 bins of the kinematic variable on the horizontal axis.}
    \label{fig:knunu-map}
\end{figure}

The null (SM) and alternative (BSM) predictions have also been calculated using the \EOS software~\cite{EOSAuthors:2021xpv,EOS:v1.0.11} and are shown in \cref{fig:knunu-dist} .
The null joint number density is obtained by binning the MC data in a 2-dimensional histogram of the reconstruction variable $q_{rec}^2$ against the kinematic variable $q^2$. This is shown in \cref{fig:knunu-map} .

\subsubsection{Full statistical model}
\label{sec:stat-model-K}
To build the posterior according to the statistical model described in \cref{sec:implementation}, we collect all parameters of our likelihood and their corresponding priors.

The theoretical parameters include the WET parameters and the hadronic parameters.
The WET parameters correspond to the three independent linear combinations of Wilson coefficients
that enter the theoretical description of $B\to K \nu \bar \nu$ decays; see \cref{eq:width}.
These are $|C_{VL} + C_{VR}|$, $|C_{SL} + C_{SR}|$, and $|C_{TL}|$.
While the Wilson coefficients are -- in general -- complex-valued parameters,
we note that the overall phase of the WET Lagrangian \cref{eq:lagrangian} is not observable.
Moreover, an inspection of the differential decay rate in \cref{eq:width} shows only sensitivity to the absolute values of the three linear combinations of Wilson coefficients.
Hence, we represent each linear combination as a positive real-valued number. 
Their prior is chosen as the uncorrelated product of uniform distributions with support 
\begin{equation}
    \begin{aligned}
        & 5 < |C_{VL}+C_{VR}| < 20,\\
        & 0 < |C_{SL}+C_{SR}| < 15,\\
        & 0 < |C_{TL}| < 15.
    \end{aligned}
\end{equation}
The correlated hadronic parameters describe the $B\to K$ form factors as discussed in \cref{sec:knunu-theory}.
Their prior is a multivariate normal distribution, which is implemented as a sequence of independent univariate normal distributions, as discussed at the end of \cref{sec:implementation-reweighting}.

The experimental constraint includes one parameter per bin of the reconstruction variable, representing the statistical uncertainty of the MC yields.
The prior for these parameters are normal distributions $\mathcal{N}(1,1/\sqrt{N_b})$, where $N_b$ is the total yield in reconstruction bin $b$.
For the purpose of this proof-of-concept study, we do not account for further (systematic) sources of uncertainty.

\subsubsection{Reinterpretation results}
\begin{figure}
    \centering
    \includegraphics[width=0.5\textwidth]{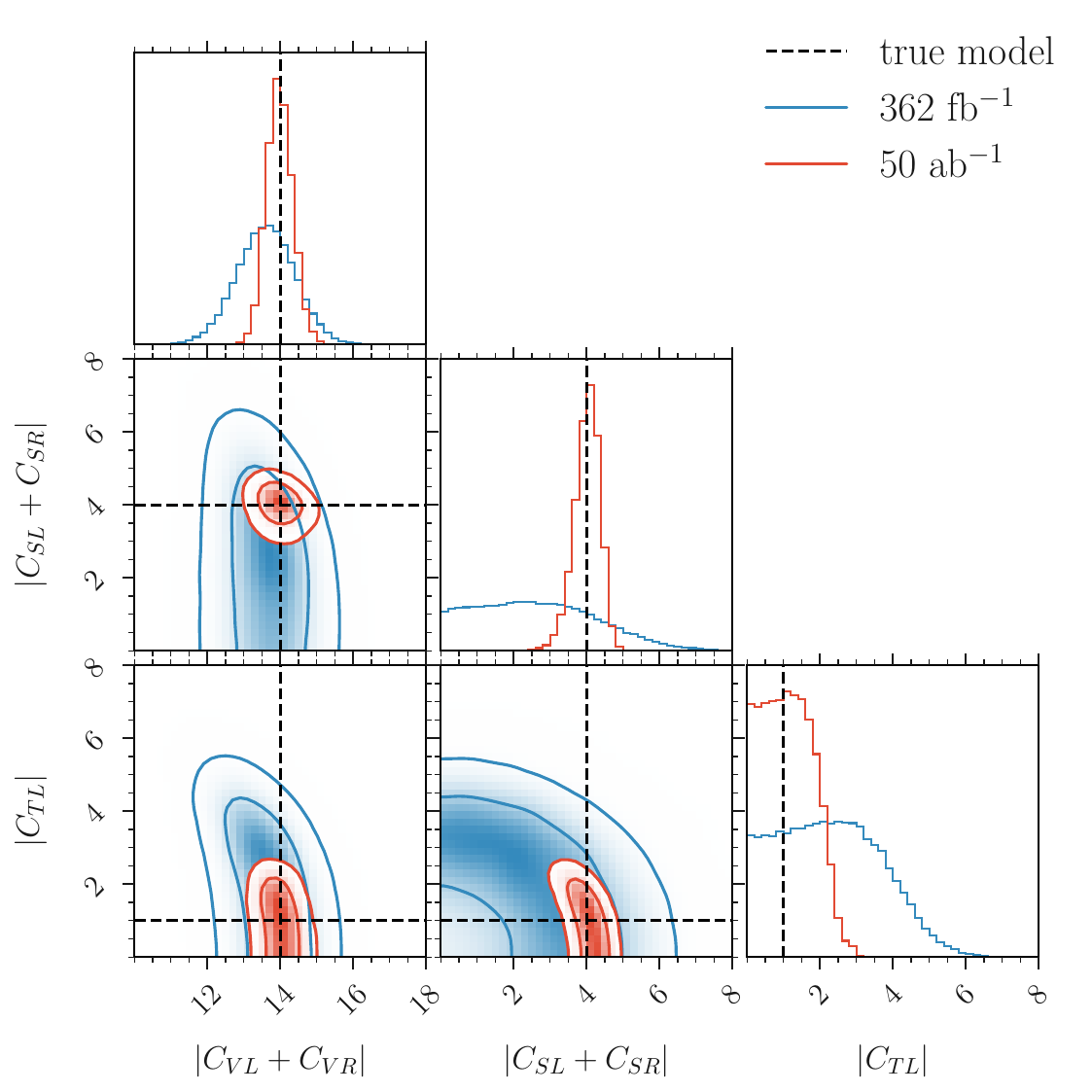}
    \caption{The marginalized posterior distributions, obtained by MCMC sampling from the $B \to K \nu \bar \nu$ likelihood. On the diagonal, we see the 1-dimensional marginal distributions of the Wilson coefficients in \cref{eq:BToKnunu-wc-sensitivity}. The contours on the 2-dimensional plots correspond to $68\%$ (inner) and $95\%$ (outer) probability. The dashed lines indicate the true underlying model (\cref{eq:WET-BSM-benchmark-Knunu}).}
    \label{fig:knunu-posterior}
\end{figure}

Having built a model-agnostic likelihood function from our toy data, we investigate the potential of our approach to constrain the Wilson coefficients.
Using MCMC sampling, we obtain the 3-dimensional marginal posterior distribution of the Wilson coefficients. 
The values at the mode of the full posterior agree with those of the benchmark point outlined in \cref{eq:WET-BSM-benchmark}. 
We show the full set of 2-dimensional marginalizations of this posterior and the resulting intervals at $68\%$ and $95\%$ probability in \cref{fig:knunu-posterior}.

We find that the marginal posterior peaks at the expected point, \cref{eq:WET-BSM-benchmark-Knunu}.
The anti-correlation of the scalar and tensorial Wilson coefficients can be seen in their marginalized 2-dimensional distribution. This behaviour is not surprising, as the tensorial and scalar terms in \cref{eq:width} peak at larger values of $q^2$, where the efficiency (\cref{eq:eff-knunu}) is low.
Moreover, the observed behaviour weakens as the statistical power of the data increases.

Overall, we see a good agreement with the expected Wilson coefficients, which acts as a closure test for our method.

\subsection{Combination of \texorpdfstring{$B \to K \nu \bar \nu$}{B->Knunubar} and \texorpdfstring{$B \to K^* \nu \bar \nu$}{B->K*nunubar}}
\label{sec:combination}

A limitation of studying solely the $B \to K \nu \bar \nu$ process is that its sensitivity to the WET Wilson coefficients is limited to the three linear combinations shown in \cref{eq:BToKnunu-wc-sensitivity}.
In this example, we showcase the power of combining data on $B \to K \nu \bar \nu$
and $B \to K^* \nu \bar \nu$ decays. These decays exhibit \emph{complementary sensitivity} to the Wilson coefficients, due to their different hadronic spin and orbital angular momentum configurations. 
For the sake of simplicity of this example, we neglect effects of additional kinematic variables in the decay chain
$B\to K^*(\to K \pi)\nu\bar\nu$, such as the helicity angle $\theta_K$ of the kaon and the $K \pi$ invariant mass. For the application of our proposed method to a real-world example, all kinematic variables should be included in the joint number density for full reinterpretability.

Moreover, this example shows from a technical perspective that our method and its implementation work also for combined \pyhf models, providing full access to the complementarity in the sensitivity.\\

\subsubsection{\texorpdfstring{$B \to K^* \nu \bar \nu$}{B->K*nubarnu} WET parametrization}
\label{sec:ksnunu-theory}

\begin{figure}[t]
    \centering
    \includegraphics[width=\linewidth]{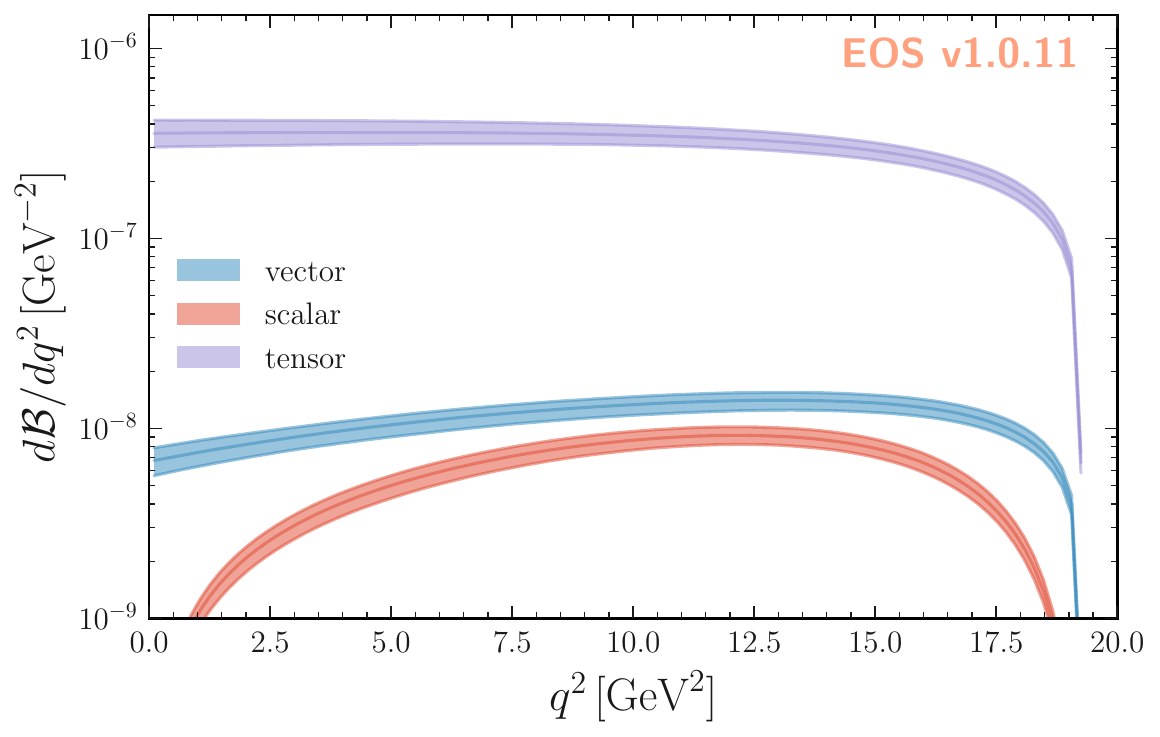}
    \caption{
        Illustration of the variety of shapes of the $B \to K^* \nu \bar \nu$ decay rate due to purely vectorial, scalar, or tensorial interactions.
        Each curve corresponds to setting a single (left-handed) non-zero Wilson coefficient in \cref{eq:widthKs} to unity while keeping all other coefficients at zero.
    }
    \label{fig:ksnunu-theory}
\end{figure}

The decay $B \to K^* \nu \bar \nu$ is governed by the same WET Lagrangian as described by
\cref{eq:lagrangian,eq:operators}.
Its differential decay rate reads~\cite{Gratrex:2015hna,Felkl:2021uxi}
\begin{widetext}
\begin{multline}
    \frac{d \Gamma}{d q^2}
        = 3 \left(\frac{4 G_\text{F}}{\sqrt{2}} \frac{\alpha}{2 \pi} \right)^2 \left|V_{t s}^* V_{t b}\right|^2
        \frac{\sqrt{\lambda_{B K^*}} q^2}{(4 \pi)^3 M_B^3}\\
        \times
          \left[|\mathcal{A}_V|^2\left|C_{\mathrm{VL}}+C_{\mathrm{VR}}\right|^2\right.
        + |\mathcal{A}_A|^2 \left|C_{\mathrm{VL}}-C_{\mathrm{VR}}\right|^2
        + |\mathcal{A}_P|^2\left|C_{\mathrm{SR}}-C_{\mathrm{SL}}\right|^2
        + \left. |\mathcal{A}_T|^2 \left|C_{\mathrm{TL}}\right|^2\right] \, ,
\label{eq:widthKs}
\end{multline}
where the reduced amplitudes multiplying the Wilson coefficients read
\begin{equation}
\begin{aligned}
    |\mathcal{A}_V|^2
        & = \frac{\lambda_{B K^*}|V(q^2)|^2}{12\left(M_B+M_{K^*}\right)^2} \, ,\quad
    &
    |\mathcal{A}_A|^2 
         &= \frac{8 M_B^2 M_{K^*}^2}{3 q^2}\left|A_{12}(q^2)\right|^2
         +\frac{\left(M_B+M_{K^*}\right)^2\left|A_1(q^2)\right|^2}{12} \, ,\\
    |\mathcal{A}_P|^2 
        & = \frac{\lambda_{B K^*}}{8\left(m_b+m_s\right)^2}\left|A_0(q^2)\right|^2 \, , \quad
    &
    |\mathcal{A}_T|^2
         &= \frac{32 M_B^2 M_{K^*}^2\left|T_{23}(q^2)\right|^2}{3\left(M_B+M_{K^*}\right)^2} +\frac{4 \lambda_{B K^*}\left|T_1(q^2)\right|^2}{3 q^2} 
     +\frac{4\left(M_B^2-M_{K^*}^2\right)^2\left|T_2(q^2)\right|^2}{3 q^2} \, .
\end{aligned}
\end{equation}
The description below \cref{eq:width} applies here as well.
\end{widetext}
In order to understand the individual contributions to the differential decay rate in \cref{eq:width} by vectorial, scalar, and tensorial operators, we provide an illustration of their relative sizes and their shapes in \cref{fig:ksnunu-theory}. This is achieved by setting their respective Wilson coefficients to unity.

One readily finds that the dependence of the observables on the Wilson coefficients is very different in \cref{eq:width,eq:widthKs}, respectively.
Compared to $B\to K\nu\bar\nu$ decays, the differential $B\to K^*\nu\bar\nu$ decay rate
exhibits additional sensitivity to the quantities
\begin{equation}
    |C_{VL} - C_{VR}|, ~ |C_{SL} - C_{SR}|.
    \label{eq:BToKstarnunu-wc-sensitivity}
\end{equation}
As a consequence, a simultaneous analysis of both decays allows constraining a total of five real-valued (out of ten total real-valued) parameters in the $sb\nu\nu$ sector. Assuming all WET Wilson coefficients to be real-valued, this corresponds to constraining the magnitudes of all Wilson coefficients. For an illustrative example, we apply this assumption here.

The hadronic matrix elements of the WET operators in this decay are expressed in terms of seven independent form factors $V(q^2)$, $A_0(q^2)$, $A_1(q^2)$, $A_{12}(q^2)$, $T_1(q^2)$, $T_2(q^2)$ and $T_{23}(q^2)$, which are functions of the momentum transfer $q^2$.
In this work, these form factors are parametrized following the BSZ parametrization~\cite{Bharucha:2015bzk},
which is truncated at order $K=2$.
The values for the corresponding $19$ hadronic parameters arise from the Gaussian likelihood provided in reference~\cite{Gubernari:2023puw}.
Correlations between the hadronic parameters are taken into account through their covariance matrix and implemented as discussed in \cref{sec:implementation-reweighting}.

\subsubsection{Datasets}

\begin{table}[t]
    \centering
    \begin{tabular}{ cccc } 
    \hline
    Luminosity & $B \overline{\kern -0.18em B}$ events & \makecell{MC generated \textbackslash \\ reconstructed} &\makecell{Data generated \textbackslash \\ reconstructed} \\
    \hline
    $362~\text{fb}^{-1}$ & $\sim 3.87 \times 10^8$ & \makecell{$3.61\times 10^3$ \\ 846} & \makecell{ $1.05\times 10^4$ \\ $2.49\times 10^3$} \\ 
    $50~\text{ab}^{-1}$   & $\sim 5.35 \times 10^9$ & \makecell{$4.99\times 10^5$ \\ $1.16\times 10^5$} & \makecell{$1.45 \times 10^6$ \\ $3.43 \times 10^5$}\\ 
    \hline
    \end{tabular}
    \caption{The number of $B \to K^* \nu \bar \nu$ samples produced for this study, corresponding to an equivalent of $50~\text{ab}^{-1}$ integrated luminosity at the SuperKEKB collider. \textit{Generated} and \textit{reconstructed} samples correspond to the numbers prior and post efficiency correction.}
    \label{tab:samplesKs}
\end{table}

To produce the $B \to K^* \nu \bar \nu$ datasets, we adapt the same procedure as in \cref{sec:knunu-data}. 

The MC data is produced according to the SM prediction (\textit{null} hypothesis).
The number of samples is calculated by multiplying the estimated number of $B \overline{\kern -0.18em B}$ events in a $50~\text{ab}^{-1}$ Belle II dataset with the predicted SM branching fraction, $BR(B \to K \nu \bar \nu) \approx 9.34 \times 10^{-6}$~\cite{EOSAuthors:2021xpv,EOS:v1.0.11}.

The real data is produced according to the BSM prediction of the benchmark point in \cref{eq:WET-BSM-benchmark} (\textit{alternative} hypothesis). The number of data samples is calculated by multiplying the estimated number of $B \overline{\kern -0.18em B}$ events in a $50~\text{ab}^{-1}$ Belle II dataset with the predicted BSM branching fraction, $BR(B \to K^* \nu \bar \nu) \approx 2.72 \times 10^{-5}$~\cite{EOSAuthors:2021xpv,EOS:v1.0.11}.

We list the number of samples in \cref{tab:samplesKs}.
We produce MC samples of the decay's probability distribution for both the null and the alternative hypothesis using the \EOS software in version 1.0.11~\cite{EOS:v1.0.11}.

The efficiency map in this case is chosen to be 
\begin{equation}
    \varepsilon(q^2) = 0.3\left(1-0.08\exp\left(- 2.5 ~ q^2/M_B^2\right) \right),
\end{equation}
which is an approximate expectation for an inclusive $B \to K^* \nu \bar \nu$ analysis. 
For the sake of simplicity, we assume that the efficiency is independent of the helicity angle $\theta_K$ and the $K\pi$ invariant mass.

We choose 10 bins in the reconstruction variable ($q_{rec}^2$) and find that 25 bins in the kinematic variable ($q^2$) provide a sufficient accuracy, using the procedure described in \cref{app:kinematic-binning,sec:knunu-data}.

\begin{figure}[ht]
    \centering
    \includegraphics[width=\linewidth]{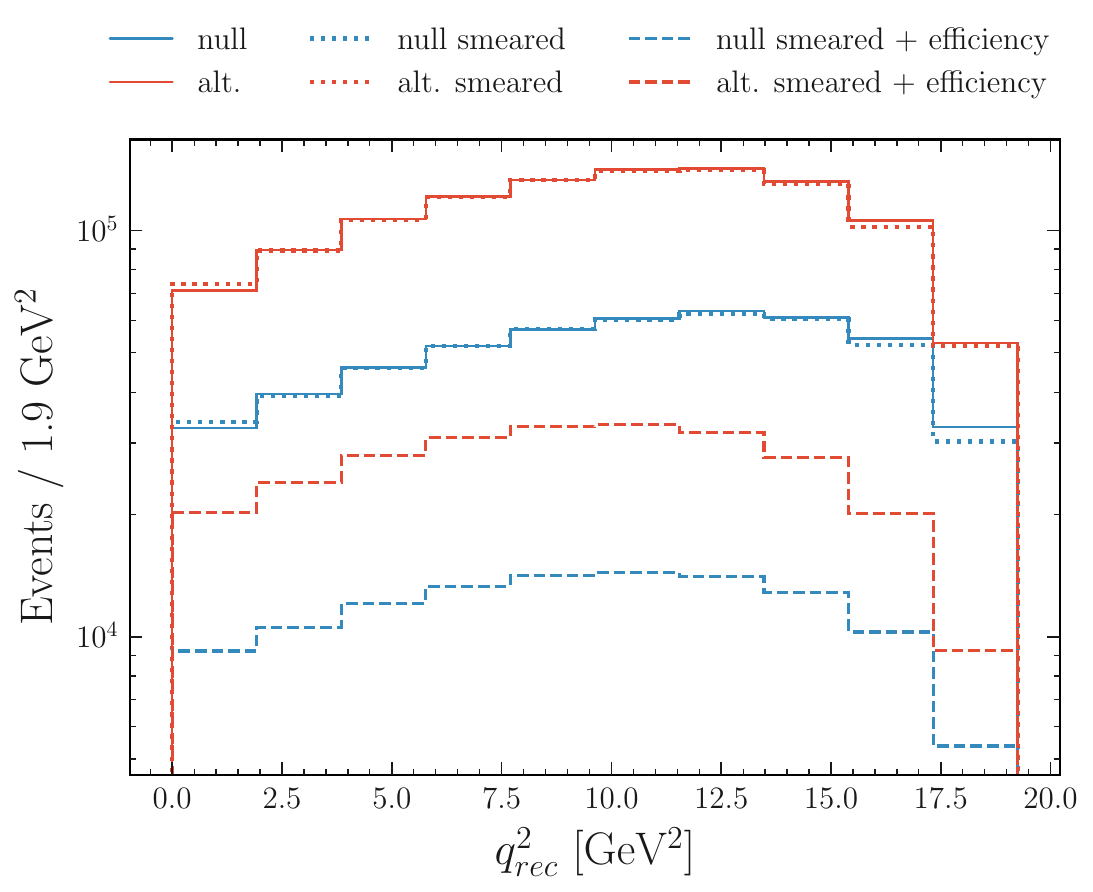}
    \caption{Both the null/SM (blue lines) and alternative/BSM (red lines) $B \to K^* \nu \bar \nu$ datasets, according to the pure theoretical prediction, after detector resolution smearing and efficiency correction.}
    \label{fig:kstarnunu-data-sets}
\end{figure}

Both datasets, according to the null (SM) and alternative (BSM) hypothesis, and their corresponding changes after detector resolution smearing and efficiency correction are shown in \cref{fig:kstarnunu-data-sets}.

\subsubsection{Full statistical model}

In order to derive the statistical model encompassing $B \to K \nu \bar \nu$ and $B \to K^* \nu \bar \nu$, we construct individual posteriors for each channel following the methodology outlined in \cref{sec:stat-model-K}. The combined posterior arises from their product. The WET theory parameters are the only parameters shared by the individual posteriors.

These parameters correspond to the five magnitudes of Wilson coefficients that enter the theoretical description of $B\to K \nu \bar \nu$ and $B\to K^* \nu \bar \nu$ decays; see \cref{eq:width,eq:widthKs}. They are $C_{VL}$, $C_{VR}$, $C_{SL}$, $C_{SR}$, and $C_{TL}$. Their prior is chosen as the uncorrelated product of uniform priors with support 
\begin{equation}
    \begin{aligned}
        &5 < |C_{VL}| < 15,\\
        &0 < |C_{VR}| < 10,\\
        &0 < |C_{SL}| < 10,\\
        &0 < |C_{SR}| < 10,\\
        &0 < |C_{TL}| < 10.
    \end{aligned}
\end{equation}

The hadronic parameters, describing the $B\to K$ and $B\to K^*$ form factors are discussed in \cref{sec:knunu-theory,sec:ksnunu-theory}. Their prior is a multivariate normal distribution, which is implemented as a sequence of independent univariate normal distributions, as discussed at the end of \cref{sec:implementation-reweighting}.

In the context of \HistFactory models, the combined likelihood of $B \to K \nu \bar \nu$ and $B \to K^* \nu \bar \nu$ is a combination on the \textit{channel} level, as discussed in \cref{sec:implementation}. One custom modifier is added to each channel, which are functions of the same WET parameters, but different hadronic parameters.

\subsubsection{Reinterpretation results}

\begin{figure}
    \centering
    \includegraphics[width=0.5\textwidth]{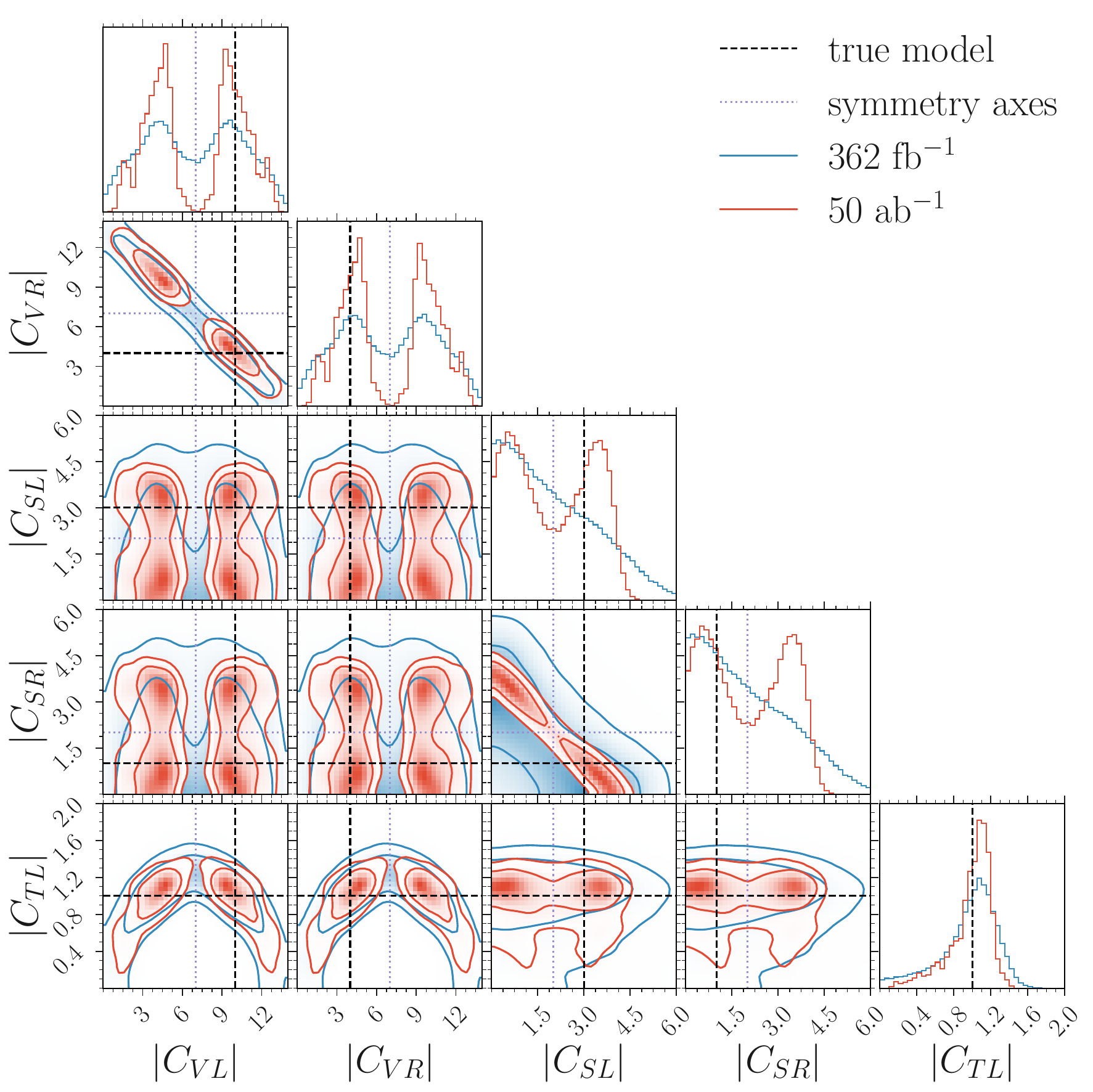}
    \caption{
    The marginal posterior distributions, obtained by MCMC sampling from the combined $B \to K \nu \bar \nu$ and $B \to K^* \nu \bar \nu$ likelihood. On the diagonal, we see the 1-dimensional marginal distributions of the Wilson coefficients appearing in \cref{eq:width,eq:widthKs}. The contours on the 2-dimensional plots correspond to $68\%$ (inner) and $95\%$ (outer) probability. The dashed lines indicate the true underlying model. The dotted lines indicate the symmetry axes of the global likelihood.}
    \label{fig:combination-posterior}
\end{figure}

From the constructed model-agnostic likelihood function, we investigate the power of constraining the full set of Wilson coefficients appearing in \cref{eq:width,eq:widthKs}, under the assumption that they are real-valued.
The decay rates in \cref{eq:width,eq:widthKs} exhibit two discrete symmetries; one under the exchange ${C_{VL} \leftrightarrow C_{VR}}$ and another under the exchange ${C_{SL} \leftrightarrow C_{SR}}$. The combination of both symmetries leads to a
four-fold ambiguity for the extraction of the Wilson coefficients from data and therefore a multimodal posterior density.
To avoid computational issues in sampling from the posterior, we select one of the four fully equivalent modes for
sampling. We do so by imposing the additional constraints $C_{VL} > C_{VR}$ and $C_{SL} > C_{SR}$.
We use MCMC sampling and initialize the chains with the mode of the full posterior.
The values at the mode of the full posterior align with those of the benchmark point outlined in \cref{eq:WET-BSM-benchmark}. 
To obtain the full multimodal posterior, we restore the original symmetry manually.
From the symmetrized samples, we obtain the 5-dimensional marginal posterior distribution of the Wilson coefficients.
We illustrate it in \cref{fig:combination-posterior} by showing the full set of 1- and 2-dimensional marginalizations and the resulting regions at $68\%$ and $95\%$ probability.

Most significantly, we see that it is now possible to probe the magnitudes of all 5 Wilson coefficients by combining results for $B \to K \nu \bar \nu$ and $B \to K^* \nu \bar \nu$. In terms of accuracy, the benefits of this combination are especially visible for $|C_{TL}|$, compared to the $B \to K \nu \bar \nu$ result in \cref{fig:knunu-posterior}.
The improvement in precision of the $50~\text{ab}^{-1}$ over the $362~\text{fb}^{-1}$ datasets is also clearly visible. This is especially prominent in the scalar sector.
For the smaller dataset, the peaks overlap such that the modes are not clearly separated. In addition, the fact that we are sampling the magnitudes of the Wilson coefficients causes an asymmetry in the scalar distributions.
A tail of $|C_{TL}|$ towards lower values is present, as in the previous example in \cref{fig:knunu-posterior}.

As in the previous example, this study serves as a further successful closure-test for our reinterpretation method. Furthermore, it shows how efficiently combinations of measurements can be performed with this method. 

\subsection{The necessity for reinterpretation}
\label{sec:necessity}
The availability of open-datasets for most particle physics results is currently very limited, although improving, thanks to the popularity of novel statistical approaches such as \HistFactory and tools such as \pyhf. These current limitations regularly hinder theorists to fully interpret existing experimental results in their BSM analyses. In particular, BSM changes to the distribution of the reconstruction variable are routinely neglected. In fact, the most common approach in a BSM analysis is to constrain the ratio of BSM prediction over SM prediction from branching ratio measurements or upper limits. This approach is only valid if the BSM changes to the shape of the distribution of the kinematic variable can be accounted for by a systematic experimental uncertainty in the reconstruction space. As we show in the following, this does not hold for measurements of the branching ratio of $B\to K\nu\bar\nu$.

To illustrate the issue, we compare our results in \cref{sec:knunu} based on simulated data with those obtained from a naive rescaling of the branching fraction. In the language of the presented reinterpretation method, the latter corresponds to using only a single bin in the kinematic d.o.f., covering the full kinematic range. This further translates to a single weight applied to all bins of the reconstruction variable, corresponding to the ratio of the alternative to the null prediction integrated over the full kinematic range. We therefore construct a further ``naive'' $B \to K \nu \bar \nu$ posterior, which deviates from the setup in \cref{sec:knunu} only by using a single bin in the kinematic range.

\begin{figure}
    \centering
    \includegraphics[width=0.5\textwidth]{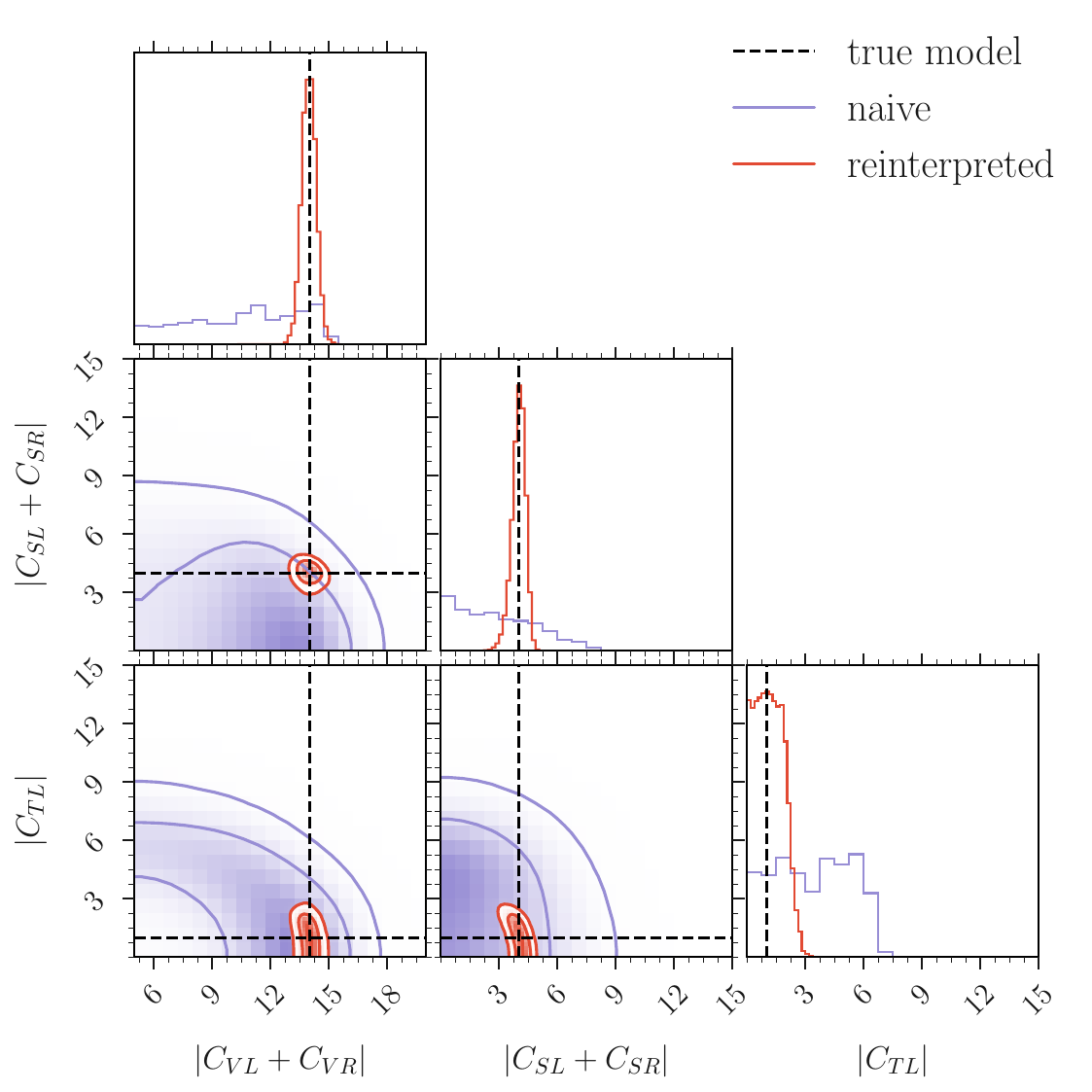}
    \caption{The comparison of the posterior distribution resulting from a model with only one bin in the kinematic d.o.f. to the proposed reinterpretation method, respecting shape changes in the kinematic distribution (see \cref{fig:knunu-posterior}).}
    \label{fig:knunu-posterior-compare}
\end{figure}

After sampling from this ``naive'' posterior, we compare the marginal distributions for the Wilson coefficients to those presented in \cref{fig:knunu-posterior}. This comparison is shown in \cref{fig:knunu-posterior-compare}.
We find a striking difference in the overall shape of the distributions and the central intervals at $68\%$ ($95\%$) probability.
Clearly, the ``naive'' procedure fails to validate, yielding large deviations from the benchmark point in \cref{eq:WET-BSM-benchmark-Knunu} in all three sectors.
Our results illustrate that our approach is essential for a faithful reinterpretation of the experimental results of $B\to K\nu\bar\nu$. 

We want to emphasize that our method provides a means to ensure an \emph{accurate} interpretation of the existing likelihood beyond the
assumptions of the underlying signal model. This does not imply, however, that our interpretation is more \emph{precise} than
a naive BSM interpretation. Put differently,
our approach eliminates a bias introduced by using an incorrect template for the decay's kinematic distribution,
however, at the expense of potentially larger uncertainties on the theory parameters.

\section{Discussion and significance of the method}

We present a novel reinterpretation method for particle physics results, which is simple in its application and requires only minimal information in addition to published likelihoods.

Our proposed method avoids biases that are introduced in the naive reinterpretation of the data at a negligible increase of compute time.
As such, it provides most of the benefits of reinterpretation using full analysis preservation.
Therefore, this method provides good trade-off between accuracy and speed, which also has the potential to be used for improving the accuracy of global effective field theory fits to many analysis results.

To showcase the method, we apply it to a simulated dataset of the $B \to K \nu \bar \nu$ decay, inspired by the recent Belle II analysis~\cite{Belle-II:2023esi} but without resorting to using any public or private Belle II data.

Using the two examples discussed in \cref{sec:examples}, we validate our method by successfully recovering the benchmark theory point from the underlying synthetic data.
This outcome underscores the accuracy and self-consistency of our approach.
We further investigate the bias introduced by naive rescaling of the $B\to K\nu\bar\nu$ branching ratio.
For our benchmark point, we find a sizable bias when determining the WET Wilson coefficient without the application of our method. 

In conclusion, this paper illustrates the ease of applicability of and the urgent necessity for
shape-respecting reinterpretation over the traditional approach.
We hope that this work will motivate experimental collaborations and analysts to consider future reinterpretability of their results and to publish the necessary material (for further details, see \cref{app:recipe}). This will, in turn, enable the whole community to use the analysis results with accuracy.

\section{Acknowledgements}
L.G. is funded by the Deutsche Forschungsgemeinschaft (DFG, German Research Foundation) – project number 460248186 (PUNCH4NFDI).
M.H. and L.H. are supported by the Excellence Cluster ORIGINS, which is funded by the Deutsche Forschungsgemeinschaft (DFG, German Research Foundation) under Germany’s Excellence Strategy - EXC-2094-39078331.
D.v.D.~acknowledges support by the UK Science and Technology Facilities Council (grant numbers ST/V003941/1 and ST/X003167/1).

\appendix
\renewcommand{\thesection}{\Alph{section}}
\renewcommand{\thesubsection}{\Alph{section}.\arabic{subsection}}

\section{Code repository and examples}
The code is available at this \href{https://github.com/lorenzennio/redist}{repository}~\cite{redist_v1_0}. In the \texttt{examples} folder, one can find the examples described in this work.

Statistical inference is performed using \pyhf \cite{pyhf,Heinrich:2021gyp}. Theoretical predictions are obtained from \EOS \cite{EOSAuthors:2021xpv,EOS:v1.0.11}.

\section{Singular value decomposition}
\label{app:svd}
Singular value decomposition is a useful method for decorrelating a set of parameters by a unitary transformation. 

In our case, we start with a covariance matrix $C$, which is symmetric. Hence, we can always decompose it as 
$$
C = USU^H
$$
where $UU^H = \identity$. The columns of the transformation matrix $U$ are the eigenvectors of $C$.
The  eigenvalues, $s_i = S_{ii}$, are the variances in the rotated space. The standard deviations are $\sigma_i = \sqrt{s_i}$.

If we want to incorporate the variances, we can define a new transformation matrix $Z = U \sqrt{S}$. $Z$ is column-wise composed of the eigenvectors of $C$, each of which is now scaled by the corresponding standard deviation.

The \pyhf modifier parameters, $\bm{p}$, describe the contribution of each of these scaled eigenvectors. Hence, to rotate from these parameters to the standard deviation vector for the correlated parameters, $\bm{\alpha}$, we can use
$$ \bm{\sigma}_\alpha = Z \bm{p}.$$
That way, the modifier parameters are all interpretable in the same way as the usual \pyhf modifier parameters, i.e. that a $p_i = \pm 1$ corresponds to a shift of $\pm \sigma_i$ along the $i^{th}$ eigenvector direction. 

\section{A recipe for application of this reinterpretation method}
\label{app:recipe}
To assist with an easy application of this reinterpretation method to any analysis, we provide a simple 4-step guide on what needs to be done to reinterpret a result from high energy physics. We focus here on the discrete approach of \cref{sec:reweighting-method-discrete}.
\begin{enumerate}
    \item \textit{Samples}. Gather your post-reconstruction samples and ensure that they contain information of all kinematic d.o.f. as well as the reconstruction variable.
    \item \textit{Null joint number density}. From these samples, build the null joint number density by simply binning samples in bins of the reconstruction variable times the kinematic d.o.f. (see \cref{app:kinematic-binning} on how to optimize the kinematic binning).
    \item \textit{Weights}. Identify your null prediction used for producing the original MC samples. Chose your alternative theoretical prediction(s) and ensure that the support of the null distribution covers the full range of the alternative distribution (this can also be done by setting an upper bound on the weights). Compute the weights as the ratio of the bin-integrated alternative to the bin-integrated null distribution (as in \cref{eq:binned-weights}).
    \item \textit{Inference}. Either making use of the code in~\cite{redist_v1_0} or by implementing \cref{eq:reweight_discrete}, compute the expected yields, given the alternative prediction, making use of the joint number density and the computed weights. Using either \pyhf~\cite{pyhf,Heinrich:2021gyp} or alternative tools, statistical inference can be used to compute results for the alternative theory. 
\end{enumerate}

\subsection{Kinematic binning}
\label{app:kinematic-binning}
To obtain suggestions on the number of bins to use for the kinematic variable(s), one can follow a similar procedure, as already mentioned in \cref{sec:knunu-data}.

For a large set of models, covering your parameter space as thoroughly as possible, compute the expected yields for a finer and finer binning in the kinematic d.o.f. (a new joint number density and new weights need to be computed every time). At each step, compute the difference to the results of the previous step and stop when reaching a pre-defined convergence condition. The maximum number of bins over all the looped models should give a good estimate on the number of kinematic bins to use.

\newpage

\bibliography{references}
\bibliographystyle{apsrev4-1}

\end{document}